\documentclass[twocolumn,showpacs]{revtex4}
\usepackage[xdvi]{graphicx}%
\usepackage{dcolumn}
\usepackage{epsfig}

\begin{document}
\title{High-temperature expansion for steady nonequilibrium states in
driven lattice gases}
\author{Raphael Lefevere and Hal Tasaki}
\affiliation
{Theoretical Physics Institute, Universite de Louvain,1348 Louvain-la-Neuve, Belgium, Department of Physics, Gakushuin University,
Mejiro, Toshima-ku, Tokyo 171, Japan}

\date{\today}

\begin{abstract}
We develop a controlled high-temperature expansion for nonequilibrium steady states of the driven lattice gas.
We represent the steady state as $P(\eta)\propto e^{-H(\eta)-\Psi(\eta)}$, and  evaluate the lowest order contribution to the nonequilibrium effective interaction $\Psi(\eta)$.
We see that, in dimensions $d\ge2$, all models with nonsingular transition rates have the same $\Psi(\eta)$, which consists of a three-body effective interaction decaying like $1/r^{2d+1}$.
The models with the Metropolis rule show exceptional behavior.
\end{abstract}

\maketitle

To construct statistical mechanics that apply to  
nonequilibrium states 
has been among the most challenging open problems in theoretical physics. 
So far it is not known how the desired 
probability measures for nonequilibrium states look like,
or even whether the measures can be written in 
universal and compact forms.

In a standard approach to nonequilibrium systems, one studies a concrete many-body problem with a given microscopic dynamics, hoping to get some hints for  possible universal statistical descriptions.
Among the most well studied models is the {\em driven lattice gas}\/ \cite{KLS,S,SZ}, in which particles on a lattice are driven into one direction by an external electric field.
For us, a major issue concerning that model is to elucidate universal aspects of the nonequilibrium steady state.

Here, we develop a high-temperature expansion for the nonequilibrium steady state of the driven lattice gas.
We represent the steady state as in (\ref{e:PeHP}) with the {\em nonequilibrium effective interaction}\/ $\Psi(\eta)$.
We explicitly compute $\Psi(\eta)$ in the lowest order of the expansion, without making any unjustified assumptions (such as the truncation of higher order correlations).
Although there were some attempts to develop expansions in the driven lattice gas \cite{ZWLV,GLMS,SZ}, our approach to evaluate the effective interaction explicitly seems to be novel.
Recently field theoretic perturbation method was developed for one-dimensional soft core lattice gas models driven by boundary conditions \cite{vWR}.

We will find that, in the lowest order ($O(\beta^2)$ for $d\ge2$, and $O(\beta^2)$ or $O(\beta^3)$ for $d=1$), the effective interaction can have only two-body and three-body interactions as in (\ref{e:Psi}).
Interestingly, these effective interactions generally are non-local and exhibit power law decay even though the Hamiltonian is completely short ranged.
Such observations may be of fundamental importance when one seeks for a universal description of nonequilibrium steady states.

When the transition rate function $\phi(h)$ is non-singular and the dimension is higher than one, our result of the expansion is rather satisfactory, suggesting a universal behavior of the nonequilibrium steady states.
We see that in the lowest order, the nonequilibrium effective interaction $\Psi(\eta)$ consists only of an effective three-body interaction $\psi^{(3)}_{x,y,z}$ and is completely independent of the transition rule and the density.
The effective interaction decays as in (\ref{e:psi3}) and is summable, i.e., $\sum_{y,z}|\psi^{(3)}_{x,y,z}|<\infty$ in the infinite volume limit.

The models with the Metropolis update rule, however, show quantitatively different behavior.
This suggests that the choice of transition rule can be a serious problem in driven nonequilibrium systems, as was pointed out in \cite{vWR,Tasaki}.
Note that the Metropolis rule has been used widely in numerical computations of the driven lattice gases.

\paragraph*{Model:}
Let $\Lambda\subset{\bf Z}^d$ be the $d$-dimensional $L\times\cdots\times L$ hypercubic lattice with periodic boundary conditions.
We denote the sites as $x=(x_1,\ldots,x_d)\in\Lambda$.
A configuration of the system is described by $\eta=(\eta_x)_{x\in\Lambda}$, which is a collection of occupation variables $\eta_x$.
We set $\eta_x=1$  if $x$  is occupied by a particle and $\eta_x=0$ if $x$ is empty.
Throughout the present paper, we fix the lattice size $L$ and the particle number $N=\sum_{x\in\Lambda}\eta_x$.
The interaction between the particles is described by the Ising Hamiltonian $H(\eta)=-(J/2)\sum_{\langle x,y\rangle}\eta_x\eta_y$.  
When we write $\sum_{\langle x,y\rangle}$, we sum over all pairs of sites with $|x-y|=1$, double-counting $x,y$ and $y,x$.

The stochastic dynamics of the driven lattice gas is determined by specifying the transition rates.
For neighboring $x,y$, we set (as usual)
\begin{equation}
c(x\to y;\eta)=\delta^{(\eta)}_{x\to y}\,
\phi(H(\eta^{x,y})-H(\eta)+E(x_1-y_1)),
\label{e:cp}
\end{equation}
which is the rate for the particle to hope from $x$ to $y$ in the configuration $\eta$.
Here, $\delta^{(\eta)}_{x\to y}=\eta_x(1-\eta_y)$ is 1 only when the hop is possible, $\eta^{x,y}$ is the configuration obtained by switching $\eta_x$ and $\eta_y$ in the original configuration $\eta$, and $E$ is the electric field in the $x_1$ direction.
(We have absorbed the inverse temperature $\beta$ into $J$ and $E$.)
We use the periodic boundary conditions so that $x_1-y_1=0,\pm1$.
The function $\phi(h)$ satisfies $\phi(h)=e^{-h}\phi(-h)$ so that the {\em local detailed balance condition}
\begin{equation}
c(x\to y;\eta)=e^{H(\eta)-H(\eta^{x,y})-E(x_1-y_1)}c(y\to x;\eta^{x,y})
\label{e:LDB}
\end{equation}
holds.
We also require $\phi(h)=1+O(h)$.
Then the above condition implies $\phi'(0)=-1/2$ provided that $\phi(h)$ is differentiable at $h=0$.
The standard choices are i)~the {\em exponential rule}\/ with $\phi(h)=e^{-h/2}$, ii)~{\em the heat bath (or Kawasaki) rule}\/ with $\phi(h)=2/(1+e^h)$, and iii)~the {\em Metropolis rule}\/ with $\phi(h)=1$ for $h\le0$ and $\phi(h)=e^{-h}$ for $h\ge0$.

\paragraph*{Steady state:}
The steady state distribution $P(\eta)$ is the unique solution of
\begin{equation}
\sum_{\langle x,y\rangle}
-P(\eta)\,c(x\to y;\eta)+P(\eta^{x,y})\,c(y\to x;\eta^{x,y})
=0,
\label{e:PcPc}
\end{equation}
or, equivalently,
\begin{equation}
\sum_{\langle x,y\rangle}
c(x\to y;\eta)-\frac{P(\eta^{x,y})}{P(\eta)}\,c(y\to x;\eta^{x,y})
=0,
\label{e:cPPc}
\end{equation}
for any $\eta$ (with the fixed $N=\sum_x\eta_x$) .
Let us write the steady state distribution as
\begin{equation}
P(\eta)=\frac{1}{Z(J,E)}e^{-H(\eta)-\Psi(\eta)},
\label{e:PeHP}
\end{equation}
where $\Psi(\eta)$ is the ``effective interaction'' which represent a nonequilibrium correction to the measure.
Then (\ref{e:cPPc}) becomes
\begin{eqnarray}
&&\sum_{\langle x,y\rangle}c(y\to x;\eta^{x,y})
\,e^{H(\eta)-H(\eta^{x,y})}
\{1-e^{\Psi(\eta)-\Psi(\eta^{x,y})}\}
\nonumber\\
&&=\sum_{\langle x,y\rangle}c(y\to x;\eta^{x,y})
\,e^{H(\eta)-H(\eta^{x,y})}
-c(x\to y;\eta),
\label{e:basic}
\end{eqnarray}
which is the starting point of our perturbation theory.

Since we have $\Psi(\eta)=0$ when $E=0$ {\em or}\/ $J=0$, we shall develop an expansion in $E$ and $J$, and (here) find the lowest order contribution to the effective interaction $\Psi(\eta)$.
(We can as well think that the inverse temperature appears in the model as $J=\beta\tilde{J}$ and $E=\beta\tilde{E}$, and one expands in $\beta$.)
We first note that the left-hand side of (\ref{e:basic}) is written as $\{1+O(E)+O(J)\}\,\Delta\Psi(\eta)$, where
\begin{equation}
\Delta\Psi(\eta)=
\sum_{\langle x,y\rangle}\delta^{(\eta)}_{x\to y}\{\Psi(\eta^{x,y})
-\Psi(\eta)\}.
\label{e:LapP}
\end{equation}

Let us denote the right-hand side of (\ref{e:basic}) as $-\tilde{Q}(\eta)$.
From (\ref{e:LDB}), we have
\begin{equation}
\tilde{Q}(\eta)=
\sum_{\langle x,y\rangle}
(1-e^{E(x_1-y_1)})\,c(x\to y;\eta).
\label{e:tQ}
\end{equation}
We first assume that $\phi(h)$ is differentiable at $h=0$.
By substituting (\ref{e:cp}) into (\ref{e:tQ}) and expanding in $E$ and $J$ to the second order, we get $\tilde{Q}(\eta)\simeq Q_{\rm gen}(\eta)$ with 
\begin{equation}
Q_{\rm gen}(\eta)=
\frac{E}{2}\sum_{\langle x,y\rangle}
\delta^{(\eta)}_{x\to y}\,(y_1-x_1)\,\{H(\eta)-H(\eta^{x,y})\},
\label{e:Qgen}
\end{equation}
where the result is independent of the specific choice of $\phi(h)$.
For the Metropolis rule, which has non-differentiable $\phi(h)$, the result is different.
We get $\tilde{Q}(\eta)\simeq Q_{\rm MP}(\eta)$ with 
\begin{equation}
Q_{\rm MP}(\eta)=
-E\mathop{\sum_{\langle x,y\rangle}}_{x_1>y_1}
\delta^{(\eta)}_{x\to y}\,\{H(\eta)-H(\eta^{x,y})\},
\label{e:QMP}
\end{equation}
provided that $E>(2d-1)|J|$.
The formula is much more complicated for $|E|\le2(d-1)|J|$.

We have thus found that the lowest order contribution to the effective interaction $\Psi(\eta)$ is determined by the Poisson-like equation
\begin{equation}
\Delta\Psi(\eta)=-Q(\eta),
\label{e:DPQ}
\end{equation}
where we set $Q(\eta)=Q_{\rm gen}(\eta)$ or $Q(\eta)=Q_{\rm MP}(\eta)$ depending on the transition rules.

\paragraph*{Decomposition:}
The equation (\ref{e:DPQ}) looks intractable since it is formulated in the huge space of all possible configurations $\eta$.
Fortunately, there are beautiful decomposition properties which enable us to determine the solution rather easily.

We start from the Laplacian (\ref{e:LapP}).
For $A\subset\Lambda$, let $\eta^A=\prod_{x\in A}\eta_x$.
We expand $\Psi(\eta)$ as $\Psi(\eta)=\sum_{A\subset\Lambda}\psi_A\,\eta^A$ with real coefficients $\psi_A$.
Note that an arbitrary function of $\eta$ can be expanded in this way.
From the translation invariance and the particle number conservation, we can assume $\psi_A=0$ when $|A|=1$.
(We denote by $|A|$ the number of sites in $A$.)
Now note that $(\eta^{x,y})^A=\eta^{A_{x,y}}$ for any $\eta$ and $A\subset\Lambda$, where $A_{x,y}$ is obtained from $A$ by switching the occupation status at $x$ and $y$.
(If, for example, $x\in A$ and $y\not\in A$ then $A_{x,y}=A\backslash\{x\}\cup\{y\}$.)
Then from the linearity of the Laplacian (\ref{e:LapP}), we see that
\begin{eqnarray}
\Delta\Psi(\eta)&=&
\sum_{\langle x,y\rangle}\sum_A
\psi_A\,\delta^{(\eta)}_{x\to y}\,(\eta^{A_{x,y}}-\eta^A)
\nonumber\\
&=&
\sum_{\langle x,y\rangle}\sum_A
(\psi_{A_{x,y}}-\psi_A)\,\delta^{(\eta)}_{x\to y}\,\eta^A
\nonumber\\
&=&
\sum_A(\Delta\psi_A)\,\eta^A,
\label{e:Lapdec}
\end{eqnarray}
where
\begin{equation}
\Delta\psi_A=\mathop{\sum_{\langle x,y\rangle}}_{x\in A,\ y\not\in A}
\{\psi_{A\backslash\{x\}\cup\{y\}}-\psi_A\}.
\label{e:DPA}
\end{equation}

We then treat the lowest order charges (\ref{e:Qgen}) and (\ref{e:QMP}).
Note that $\delta^{(\eta)}_{x\to y}\,\{H(\eta)-H(\eta^{x,y})\}=\eta_x(1-\eta_y)\,J\{\sum_{z\ (|z-x|=1,z\ne y)}\eta_z-\sum_{z\ (|z-y|=1,z\ne x)}\eta_z\}$.
Since the charge $Q(\eta)$ is a linear combination of these terms, it can be expanded as 
\begin{equation}
Q(\eta)=
\sum_{x,y\in\Lambda}q^{(2)}_{x,y}\,\eta_x\eta_y
+\sum_{x,y,z\in\Lambda}q^{(3)}_{x,y,z}\,\eta_x\eta_y\eta_z,
\label{e:Qdec}
\end{equation}
with suitable coefficients $q^{(2)}_{x,y}$ and $q^{(3)}_{x,y,z}$.
Note that when evaluating the ``charges'' $q^{(2)}_{x,y}$, $q^{(3)}_{x,y,z}$, we only need to consider configurations $\eta$ with two or three particles.
This means that the calculations are (in principle) elementary.
In what follows we only present the results of charge calculations, leaving details to \cite{LT}.

By substituting these decompositions into the  Poisson equation (\ref{e:DPQ}) for $\Psi(\eta)$, we find that $\psi_A=0$ unless $|A|=2$ or 3, and $\psi_A$ with $|A|=2$, $3$ are determined by the (tractable) Poisson equations
\begin{equation}
\Delta\psi^{(2)}_{x,y}=-q^{(2)}_{x,y},
\quad
\Delta\psi^{(3)}_{x,y,z}=-q^{(3)}_{x,y,z}.
\label{e:p23}
\end{equation}
Here Laplacian $\Delta$ is defined by (\ref{e:DPA}) by identifying $\{x,y\}$ or $\{x,y,z\}$ as the subset $A$.
Equivalently they may be regarded as the standard Laplacian \cite{Laplacian} on the lattices $\Lambda_2=\{(x,y)\,|\,x,y\in\Lambda,\,x\ne y\}\subset{\bf Z}^{2d}$  or $\Lambda_3=\{(x,y,z)\,|\,x,y,z\in\Lambda,\,x\ne y,\,y\ne z,\,z\ne x\}\subset{\bf Z}^{3d}$.
Two sites in $\Lambda_2$ ({\em resp}\/. $\Lambda_3$) are regarded to be neighboring when their Euclidean distance in ${\bf Z}^{2d}$ ({\em resp}\/. ${\bf Z}^{3d}$) is equal to 1.

The effective interaction $\Psi(\eta)$ is exactly written as
\begin{equation}
\Psi(\eta)=\sum_{x,y\in\Lambda}\psi^{(2)}_{x,y}\,\eta_x\eta_y
+\sum_{x,y,z\in\Lambda}\psi^{(3)}_{x,y,z}\,\eta_x\eta_y\eta_z,
\label{e:Psi}
\end{equation}
in the lowest order.
It is remarkable that the effective interaction (at least in this order) is independent of the particle number.
We stress that we are {\em not}\/ expanding in the density.
We shall now determine $\psi^{(2)}_{x,y}$ and $\psi^{(3)}_{x,y,z}$ by solving the Poisson equations (\ref{e:p23}).

\paragraph*{General models in $d\ge2$:}
We first concentrate on general models with $\phi(h)$ being differentiable at $h=0$.
As for the two-body charge, we find that  $q^{(2)}_{x,y}=0$ for all $x$, $y$.
This means $\psi^{(2)}_{x,y}=0$.
{\em We have no two-body correction to the measure in the lowest order}\/.

The three-body charge  $q^{(3)}_{x,y,z}$, on the other hand, is nonvanishing.
Let $e_1=(1,0,\ldots,0)$, and $\cal U'$ be the set of the $2(d-1)$ unit vectors of ${\bf Z}^d$ other than $\pm e_1$.
One has $q^{(3)}_{x,x\pm e_1,x+\delta}=\pm JE$ for any $x\in\Lambda$ and any $\delta\in\cal U'$.
Other nonvanishing $q^{(3)}_{x,y,z}$ are determined by requiring it to be symmetric in $x$, $y$, and $z$.
The rest are vanishing.
To get a feeling for the behavior of the solution $\psi^{(3)}_{x,y,z}$ of the Poisson equation (\ref{e:p23}), we make an orthogonal transformation and introduce a new coordinate by $X=(2x-y-z)/\sqrt{6}$, $Y=(y-z)/\sqrt{2}$, and $Z=(x+y+z)/\sqrt{3}$.
Apparently the charge distribution  depends only on $X$ and $Y$, and so does the solution 
 $\psi^{(3)}_{x,y,z}$.
 By denoting  $\psi^{(3)}_{x,y,z}=\varphi^{(3)}_{X,Y}$ , we get $\Delta \varphi^{(3)}_{X,Y}=-\tilde{q}^{(3)}_{X,Y}$, which is the Poisson equation in the $2d$-dimensional space.
 Here the charge distribution $\tilde{q}^{(3)}_{X,Y}$ is still complicated, but when projected onto the $(X_1,Y_1)$ plane (where $X_1$ and $Y_1$ are the first coordinates of 
 $X$ and $Y$, respectively), we see that the plus charges are located at $(\sqrt{2/3},0)$, $(-1/\sqrt{6},\pm1/\sqrt{2})$ and minus charges at $(-\sqrt{2/3},0)$, $(1/\sqrt{6},\pm1/\sqrt{2})$ .
Close to the origin of the $2d$ dimensional space, there is a hexa-pole parallel to the $(X_1,Y_1)$ plane.
Roughly speaking the corresponding solution $\varphi^{(3)}_{X,Y}$ behaves like $\{1/r^{2d-2}\}'''\sim1/r^{2d+1}$.
This estimate can be made into a reliable asymptotic estimate \cite{LT} by going back to the original $3d$-dimensional lattice, and treating properly the ``mirror charge'' induced at the defect sites.  
The result is
\begin{equation}
\psi^{(3)}_{x,y,z}\sim cJE\frac{U_1V_1W_1}{(|U|^2+|V|^2+|W|^2)^{d+2}},
\label{e:psi3}
\end{equation}
where $c$ is a constant depending only on the dimension, and we have set $U=2x-y-z$, $V=2y-z-x$, and $W=2z-x-y$.
We note that quantity in the denominator can be written as $|U|^2+|V|^2+|W|^2=2\{|x-y|^2+|y-z|^2+|z-x|^2\}$.
It is notable that the $1/r^{2d+1}$ decay of (\ref{e:psi3}) implies $\sum_{y,z}|\psi^{(3)}_{x,y,z}|\le{\rm const.}$ with an $L$ independent constant.
We recall that such a summable (effective) interaction should be regarded as ``healthy'' in the light of the well established theory of the infinite volume Gibbs state \cite{Gibbs}.

To summarize, {\em the lowest order contribution to the effective interaction $\Psi(\eta)$ is independent of the transition rule provided that $d\ge2$ and $\phi'(0)$ exists}\/.
Moreover $\Psi(\eta)$  in this order consists only of the {\em summable three-body interaction}\/  with the asymptotic behavior (\ref{e:psi3}).
The robustness of the result may suggest that we have captured a universal aspect of the nonequilibrium steady state.

It should be noted, however, that the higher order contributions to $\Psi(\eta)$ may be more complicated and may become rule-dependent.
In particular the results in \cite{Tasaki} suggest that the two-body effective interaction $\psi^{(2)}_{x,y}$ is nonvanishing and exhibits a $1/r^d$ power law decay as in (\ref{e:psi}) in the order of $E^2J$ for the heat-bath rule, while there is no $1/r^d$ decay for the exponential rule.

\paragraph*{Metropolis rule in $d\ge2$:}
In models with the Metropolis rule (with $E>(2d-1)|J|$), the situation becomes completely different.
By the translation invariance, we have $q^{(2)}_{x,y}=\tilde{q}^{(2)}_{x-y}$, where $\tilde{q}^{(2)}_z=EJ$ if $z=\pm e_1$, $\tilde{q}^{(2)}_z=-EJ$ if $z=\pm2e_1$ or $z=\pm e_1+\delta$, $\tilde{q}^{(2)}_z=2EJ$ if $z=\delta$, for any $u\in\cal U'$, and $\tilde{q}^{(2)}_z=0$ otherwise.
Again, by the translation invariance the corresponding effective interaction can be written as $\psi^{(2)}_{x,y}=\varphi^{(2)}_{x-y}$, where $\varphi^{(2)}_z$ is determined by $\Delta\varphi^{(2)}_z=-\tilde{q}^{(2)}_z$.
(Here $\Delta$ is the Laplacian \cite{Laplacian} on $\Lambda\backslash\{(0,\ldots,0)\}$.)
Since $\tilde{q}^{(2)}_z$ corresponds to the charge distribution of a quadrupole in $d$-dimension, the resulting field $\varphi^{(2)}_z$ is expected to decay like $1/|z|^d$ for large $|z|$.
Again, this can be made into a precise estimate \cite{LT}, and we have
\begin{equation}
\varphi^{(2)}_z
\simeq c'\,EJ\,
|z|^{-(d+2)}
\{(d-1){z_1}^2-\sum_{j=2}^d{z_j}^2\},
\label{e:psi}
\end{equation}
for any $d\ge2$ where $c'$ is a constants which depend only on the dimension.
This power law decaying effective interaction can be regarded as the origin of the $1/r^d$ long range two-point correlation \cite{S,SZ,ZWLV,GLMS,Tasaki} known to exist in the driven lattice gas and related models.
Note that the interaction decays so slowly that the sum $\sum_{y\in\Lambda}|\psi^{(2)}_{x,y}|$ is divergent as $L\to\infty$, which implies (and indeed proves \cite{nonGibbs}) that the infinite volume limit of the steady state is not Gibbsian.
Thus, in a stark contrast with general models with nonsingular $\phi(h)$, {\em the steady state measure for  Metropolis dynamics has a significant nonequilibrium correction in the two-body sector already in the lowest order}\/.

Curiously enough, in the lowest order of the perturbation, the behavior of the three-body effective interaction $\psi^{(3)}_{x,y,z}$ is exactly the same as the general models that we have discussed.
We still do not know if this is a mere accidental coincidence or an indication of a deeper universality in the structure of nonequilibrium steady states.

\paragraph*{Models in $d=1$:}
The models in $d=1$ show exceptional behavior, which requires extra estimates \cite{LT}.

Suppose that $\phi(h)$ is twice differentiable at $h=0$.
In this case nonvanishing contribution to $\Psi(\eta)$ is found only in the third order.
The formula (\ref{e:Qgen}) for the charge is thus no longer useful.
By going back to the original definition (\ref{e:tQ}), we get the following results.
In the two-body effective interaction $\psi^{(2)}_{x,y}$, there is a short-range correction proportional to $\{2\phi''(0)-(1/2)\}E^2J$.
The three-body effective interaction $\psi^{(3)}_{x,y,z}$ exhibits a $1/r^3$  long range correlation as in (\ref{e:psi3}), and its magnitude is proportional to $\phi''(0)\,EJ^2$.
Unlike in the models with $d\ge2$, {\em the lowest order correction is already very much rule dependent}\/.
Note in particular that $\phi''(0)=0$ in the heath bath rule.

In the Appendix~4 of \cite{KLS}, one-dimensional exactly solvable models of driven lattice gases are introduced,
where the nonequilibrium steady states coincide with the corresponding equilibrium states.
In the language of the present work, these models turn out to posses vanishing charge (\ref{e:tQ}).
 We therefore recover the known results that the models have no nonequilibrium corrections.

As for the one-dimensional model with the Metropolis rule, there is nonvanishing contribution to $\Psi(\eta)$ in the order $EJ$.
We find a short range correction to $\psi^{(2)}_{x,y}$, and no contribution to $\psi^{(3)}_{x,y,z}$.

\paragraph*{Discussions:}
We have developed a high-temperature expansion for the driven lattice gas, and evaluated the effective interaction $\Psi(\eta)$ to the lowest order.
Unfortunately, we still do not know how to evaluate higher order contributions in a systematic manner (although a brute force calculation seems always possible).
We expect to get higher body effective interactions as we go to the higher orders in the perturbation.
We note that the expansion is much easier in the models with soft-core interactions, where we can take into account the effect of the field $E$ without expanding it \cite{LT}.
It is also possible \cite{LT} to write down equations for $\Psi(\eta)$ in lattice gas models driven by chemical potential difference at the boundaries, where we expect the $1/r^{d-2}$ as in \cite{S2} to be universal \cite{vWR}.

We have observed, as in \cite{vWR,Tasaki}, that the nonequilibrium steady states sometime show qualitative update-rule dependence.
It is especially sever in one-dimension, and in the Metropolis rule in any dimensions.
Such rule dependence can be a serious problem especially when one is interested in understanding universal statistical properties of nonequilibrium steady states.
It is certainly desirable to find a criterion that determines the rule which yields physically meaningful results.

We found that $\Psi(\eta)$ in $d\ge2$ is independent of the rule provided that $\phi(h)$ is differentiable.
This strongly suggests that the results thus obtained are universal and physically meaningful \cite{Metropolis}.
As we go to the higher orders in the expansion, however, even the models with differentiable $\phi(h)$ are expected to show strong rule dependence \cite{Tasaki,LT}.
We still do not know of any way to determine the ``right'' transition rule.
It is our conjecture \cite{Tasaki,LT} that the models with the exponential transition rule are free from the $1/r^d$ long range correlation, and have chance of possessing Gibbsian infinite volume steady states.
Although there seems to be no physical reasons to prefer models with Gibbsian steady state, it would be quite interesting if such a criterion plays a role in the future development of nonequilibrium statistical mechanics.

It is a pleasure to thank Joel Lebowitz, Elliott Lieb, Shin-ichi Sasa, and Herbert Spohn for useful discussions.


\begin{thebibliography}{10}
\bibitem{Laplacian}
Let $G$ be an arbitrary discrete set and $B\subset G\times G$ be the set of ``neighboring'' sites.
Then the Laplacian $\Delta$ on $G$ is defined as $\Delta f_x=\sum_{y;(x,y)\in B}(f_y-f_x)$.
\bibitem{nonGibbs}
The rough idea of the proof \cite{LT} is as follows.
We first assume that the steady state $P(\eta)$ has a sensible infinite volume limit which is a Gibbs state with interactions being analytic in the parameters $J$ and $E$.
Then the properties of the Gibbs state allows us to construct a convergent expansion for the interaction in a finite volume, which should coincide with the present expansion.
But this is impossible since our expansion generates a non-summable effective interaction.
\bibitem{Metropolis}
This in turn suggests that results and conclusions obtained using the Metropolis rule should be reconsidered.
\end{thebibliography}
\end{document}